\input amstex
\documentstyle{amsppt}
\magnification=\magstep{1.2}
\baselineskip 16 true pt
\NoRunningHeads
\NoBlackBoxes
\language0
\pageno=1 

\define\p{\partial }

\define\Om{\Omega}

\define \Si{{\Cal S}}
\define\Pa{Painlev\'e}
\define\ob#1{\overline #1}

\topmatter
\title{The Radius of Convergence and the
Well-Posedness of The Painlev\'e Expansions of the Korteweg-de-Vries
Equation}
\endtitle
\author{Short Title: Well Posedness of Painlev\'e expansions \\ \\
Nalini Joshi and Gopala K. Srinivasan}\endauthor

\affil{School of Mathematics \\ University of New South Wales\\
Sydney NSW 2052 \\Australia\\
{\tt N.Joshi\@unsw.edu.au}}\endaffil
\subjclass{35A20, 35B30, 35Q53}\endsubjclass
\abstract{In this paper we obtain explicit lower bounds for the
radius of convergence of the Painlev\'e expansions of the Korteweg-de-Vries
equation around a movable singularity manifold $\Si$ in terms of the sup
norms of the arbitrary functions involved. We use this estimate to prove the
well-posedness of the singular Cauchy problem on $\Si$ in the
form of continuous
dependence of the meromorphic solution on the arbitrary data.}
\endabstract
\endtopmatter
\document
\subheading{\S 1 Introduction}
Given a holomorphic manifold $\Si$ described locally by
$$
\Si:  x - \psi(t) = 0,            \tag 1.1
$$
the Painlev\'e expansion of the Korteweg de~Vries (KdV) equation
$$
w_t = w_{xxx} + ww_x   \tag 1.2
$$
is a formal series solution of the of the form
$$
w(x, t) = \sum_{n=0}^{\infty} u_n(t)(x - \psi(t))^{n+\nu},    \tag 1.3
$$
where $\nu=-2$, and $u_4(t)$, $u_6(t)$ are arbitrary functions.

Such series expansions
were suggested by J.Weiss, M.Tabor and G.Carnevale [7]
as a practical test of
the Painlev\'e property for partial differential equations (PDEs),
i.e. the property that
all  solutions are single-valued around all noncharacteristic
holomorphic
given singularity manifolds. The Painlev\'e property has become a
widely used
indicator for integrability (see
[1, 3]) meaning exact solvability through an associated linear
problem. The KdV equation is well known to be integrable.
Hence it is widely believed that it should possess the Painlev\'e
property.

The convergence of the \Pa\ expansions
has been discussed in [4, 6]. However, the issue
of estimating their radii of convergence has not been
addressed. Here,
we prove a lower bound on
the radius of convergence of the Painlev\'e expansion for the KdV
equation and use it to prove the
well-posedness of the singular Cauchy problem
with respect to the arbitrary data given on $\Si$.

To describe the well-posedness result in terms of such data we define
a collective name for them:
\proclaim{Definition} The {\rm WTC data} for the KdV equation
is the set $\{\psi(t), u_4(t), u_6(t)\}$
of arbitrary functions describing the Painlev\'e
expansion (1.3).
\endproclaim
\noindent We restrict our attention to the space of
holomorphic WTC data topologized by uniform convergence on compact
sets.

Our results give a lower bound for the radius of convergence in terms
of
the sup-norm of the WTC data and, moreover, show that the
meromorphic function given by the (convergent)
Painlev\'e expansion varies continuously, in the sup-norm,
as the
WTC data are varied. Before we state our main results, we need
to recall the construction and the convergence
result for the \Pa\ expansion of the KdV
equation (see Refs [4] or [7]).
\proclaim{Theorem 1.1}
Given an analytic manifold $\Si: x - \psi (t) = 0$, with
$\psi(0) = 0$, and two arbitrary analytic functions
$$
\lim_{x\to\psi(t)}\Bigl(\frac{\p}{\p x}\Bigr)^4[w(x,t)(x-\psi(t))^2], \quad
\lim_{x\to\psi(t)}\Bigl(\frac{\p}{\p x}\Bigr)^6[w(x,t)(x-\psi(t))^2]  \tag 1.4
$$
there exists in a neighbourhood of $(0,0)$ a
meromorphic solution of the KdV equation (1.2)
of the form
$$
w(x, t) = \frac{-12}{\bigl(x-\psi(t)\bigr)^2} + h(x, t)
$$
where $h$ is holomorphic.
\endproclaim

The expansion (1.3) can be written in terms of the new variable
$X:=x-\psi(t)$ as
$$
w(x,t) = \sum_{n=0}^{\infty} u_n(t) X^{n-2}. \tag 1.5
$$
Substitution into the KdV equation shows that
the coefficients $u_n$ must satisfy the recursion relation
$$
Q(n)u_n = u'_{n-3} - (n-4)\psi'(t)u_{n-2} - \sum_{j=1}^{n-1}
(j-2)u_ju_{n-j}\quad (n>0),                            \tag 1.6
$$
where $Q(n) := (n+1)(n-4)(n-6)$.
It is easily checked that $u_4$ and $u_6$ are arbitrary and
$$
u_0 = -12,\quad u_1 = u_3 = 0,\quad u_2 = -\psi'(t),\quad u_5 =
\frac{\psi''(t)}{6}.     \tag 1.7
$$
The convergence of the series may then be established by the majorant
method
of Ref. [7] or the iteration method of Ref.[4].

The main results of this paper are stated below as Theorems
A and B.
\proclaim{Theorem A (Radius of Convergence of the Series for
$w$)}
Given  WTC data $\psi(t)$, $u_4(t)$, $u_6(t)$ analytic in the ball
$B_{2\rho+\epsilon}(0) = \{t\in \Bbb C :  |t| < 2\rho +\epsilon\}$,
let
$$
M = \sup_{|t| = 2\rho}\{1,  |\psi(t)|,
|u_4(t)|, |u_6(t)|\}. \tag 1.8
$$
The radius of convergence $R_{\rho} = R$ of the power series (1.5)
satisfies
$$
R \geq \frac{\min\{1,\rho\}}{10M}.        \tag 1.9
$$
\endproclaim
Notice that, because of the Maximum Principle, the supremum over
${|t| = 2\rho}$ is the same as the supremum over ${|t| < 2\rho}$.
To describe the well-posedness result we need to be slightly more
precise in defining our notation for domains.
If the WTC data are given on
a connected bounded open set $\Om$ in $\Bbb C$ and furthermore $\delta>0$
we
define $\Om_{\delta}$ as
$$
\Om_{\delta} = \{t\in \Om :  \text{dist}(t, \Bbb C-\Om) \geq
3\delta\}.
$$
We will always assume that $\delta$ is small enough that the set
$\Om_{\delta}$
is a connected set. Note that if $t\in\Om_{\delta}$ then the \it
closed disk
\rm of radius $2\delta$ centred at $t$ is contained in
$\Om_{\delta/3}$.
We denote by
$M_{\delta}$, $B_{\delta}$ the numbers
$$
\align
M_{\delta} &= \sup_{t\in\Om_{\delta/3}}\{1, |\psi(t)|, |u_4(t)|,
|u_6(t)|\}, \\
B_{\delta} &= \frac{10M_{\delta}}{\min\{1, \delta\}},
\endalign
$$
given by Theorem A and take
$$
\align
\Cal O_{\delta} &= \{(x,t)\in \Bbb C^2: t\in \Om_{\delta} \text{ and }
|x-\psi(t)| < B_{\delta}^{-1}\},         \\
\Cal O &= \bigcup_{\delta > 0} \Cal O_{\delta}.
\endalign
$$
\proclaim{Theorem B (Continuity as a Function of WTC Data)}
\roster
\item
Let $\bigl(\psi_j(t), u_{4, j}(t), u_{6, j}(t)\bigr)$ be a sequence of
holomorphic WTC data defined on a bounded connected
domain
$\Om \in\Bbb C$ and converging
uniformly on compact subsets to $\bigl(\psi(t), u_{4}(t), u_{6}(t)\bigr)$.
Assume $\psi_j(0) = 0$ for every $j$. Let $w_j$ be the solution
of the equation (1.2)
given by (1.5) with
WTC data $\bigl(\psi_j(t), u_{4, j}(t), u_{6, j}(t)\bigr)$.
Then there exists a polydisk
$$
P(r_1, r_2) := \{(x,t): |x|<r_1, |t|< r_2\}
\subset \Cal N,
$$
where
$\Cal N
:= \bigcap_{j}\Cal O(\psi_j, u_{4,j}, u_{6,j}) \cap \Cal O(\psi, u_4,
u_6).$
\item
The sequence of functions $(x-\psi_j(t))^2w_j$ are all defined on
a common open subdomain $G \subset \Cal N$,
 and converge uniformly on compact subsets of
$G$  to $(x-\psi(t))^2w$, where $w$ is the solution of (1.2)
with data $\bigl(\psi(t),
u_{4}(t),
u_{6}(t)$.
\endroster
\endproclaim

We prove these two theorems in sections 2 and 3 respectively.

\subheading{\S 2 Proof of Theorem A -- Estimate of Radius of
Convergence}

In this section, we prove Theorem A.
The method we use is a combination of a
majorant method and iteration on a shrinking sequence of disks.
Let $v:=X^2w$.

\demo{Proof}
We estimate the sequence of numbers $\sup_{|t|= \rho}|u_n(t)|$.
Define the sequence $R_n$ as follows:
$$
R_n = \frac{6\rho}{\pi^2}\Bigl(\frac{\pi^2}{3} -
\sum_{j=1}^n \frac{1}{j^2}\Bigr).   \tag 2.1
$$
Clearly $\rho<R_n<R_{n-1}<2\rho$ for all $n$ and $R_n\rightarrow
\rho$. Define
$$
d_n = R_{n-1} - R_{n} = \frac{6\rho}{\pi^2n^2}, \quad  A :=
\pi^2/6\rho \quad \implies \frac{1}{d_n} = An^2. \tag 2.2
$$

We consider the sequence of recurrence formulae
$$
\align
Q(n)M_n &= An^2M_{n-3} + AnMM_{n-2}
+ \sum^{n-1}_{j=1}(j-2)M_{n-j}M_j
\quad (n \geq  7), \tag 2.3 \\
Q(n)U_n &= An^2U_{n-3} + AnMU_{n-2} + \sum^{n-1}_{j=1}(j-2)U_{n-j}U_j.
 \tag 2.4
\endalign
$$
For $1 \leq i \leq 6$, choose $M_i = U_i =
\sup_{|t|=R_i}|u_i(t)|$. We determine $M_n$, $U_n$ using the equations
(2.3, 2.4).

\definition{Claim} $M_i \leq U_i$ for all $i$. \enddefinition
Proof of the claim is by induction on $n$. It is true for $1\leq i \leq 6$
by the choice of $U_i$.
$$
\align
Q(n)U_n &= An^2U_{n-3} + AnMU_{n-2}
             + \sum^{n-1}_{j=1}(j-2)U_{n-j}U_j,\quad \text{using (2.4),}  \\
        &\geq An^2M_{n-3} +  AnMM_{n-2}
          +\sum^{n-1}_{j=1}(j-2) M_{n-j}M_j = Q(n)M_n
\endalign
$$
completing the induction.

\definition{Claim} $M_n \geq \sup_{|t|=R_n} |u_n(t)|$.
\enddefinition
We prove this by induction on $n$. It is true for $0\leq i \leq 6$ by
definition. By Cauchy's estimate we get
$$
\align
\sup_{|t|=R_n} |u'_{n-3}(t)|
&\leq \bigl(\sup_{|t|= R_{n-3}}|u_{n-3}(t)|\bigr)/(R_{n-3} - R_n)
\leq \frac{M_{n-3}}{(R_{n-1} - R_n)}.\\
\therefore \sup_{|t|=R_n} |u'_{n-3}(t)| &\leq
 \frac{M_{n-3}}{d_n} = An^2M_{n-3}, \text{ and}  \\
\sup_{|t|=R_n}|\psi'(t)| &\leq
\sup_{|t|=2\rho}|\psi(t)|/(2\rho - R_n) \leq \frac{M}{2\rho-R_1} = \frac{M}{A}.
\endalign
$$
Now the recursion relation defining $M_n$ gives
$$
\align
Q(n)M_n &= \frac{M_{n-3}}{d_n} + \frac{nMM_{n-2}}{2\rho-R_1}
+ \sum_{j=1}^{n-1}(j-2)M_{n-j}M_j     \\
        &\geq \sup_{|t|=R_n} |u'_{n-3}(t)|
+ n\sup_{|t|=R_n}|\psi^{'}(t)|\sup_{|t|=R_{n-2}}|u_{n-2}|   \\
  &\phantom{XXX}+
\sum_{j=1}^{n-1}(j-2)\sup_{|t|=R_{n-j}}|u_{n-j}|\sup_{|t|=R_j}|u_j|  \\
   &\geq \sup_{|t|=R_n}\Bigl[|u_{n-3}^{'}(t)| + (n-4)|\psi^{'}(t)||u_{n-2}| +
    \sum^{n-1}_{j=1}(j-2)|u_{n-j}u_j|\Bigr] \\
        &\geq \sup_{|t|=R_n}\Big| u'_{n-3}
- (n-4)\psi^{'}(t)u_{n-2}  - \sum^{n-1}_{j=1}(j-2)u_{n-j}u_j\Big|  \\
\therefore Q(n)M_n &\geq \sup_{|t|=R_n}Q(n)|u_n(t)|
\endalign
$$
completing the induction.

\definition{Claim} If we choose $K$, $B$ such that
$$
B \geq \max\{M, 10\sqrt{M/\rho},
3\rho^{-1/3}, \pi(M/\rho^2)^{1/5}, \frac{2}{\rho}\}, \quad  K = 1/4  \tag 2.5
$$
then $U_n \leq nKB^n$ for all $n$.
\enddefinition
In the following, we will use the self-evident facts
$$
n^2/Q(n) < 3, \quad \frac{n(n-1)}{(n-4)(n-6)} \leq 14\quad \text{ for } n\geq
7.
$$
The choice of $K$, $B$ and  the values of $u_n$ for $1\leq n\leq 6$
implies that $U_n \leq nKB^n$ for $1\leq n\leq 6$. We will prove this
later.  Assume inductively that the estimate is true for all $i < n$.
Now, for $n\geq 7$, we get
$$
\align
Q(n)U_n &= An^2U_{n-3} + AnMU_{n-2}
             + \sum^{n-1}_{j=1}(j-2)U_{n-j}U_j             \\
        &\leq nKB^n\Bigl( An(n-3)/B^3 + \frac{(n-2)AM}{B^2} +
K\sum_{j=1}^n j^2(n-j)/n\Bigr)
\endalign
$$
Therefore, we get
$$
\align
U_n &\leq nKB^n\Bigl(\frac{An^2}{Q(n)B^3} +
\frac{AM}{B^2(n-4)(n-6)} + \frac{Kn(n-1)}{12(n-4)(n-6)}\Bigr) \\
        &\leq nKB^n\Bigl(\frac{3A}{B^3} + \frac{A}{B(n-4)(n-6)}
                   + \frac{7K}{6}\Bigr)\\
&\leq nKB^n\Bigl(\frac{3A}{B^3} + \frac{\pi^2}{6\rho B(n-4)(n-6)} +
\frac{1}{3}\Bigr)\\
&\leq nKB^n\Bigl(\frac{3A}{B^3} + \frac{2}{3\rho B} +
\frac{1}{3}\Bigr)
\endalign
$$

To complete the induction we need to show that
\roster
\item
$3A/B^3 + 2/(3\rho B) + 1/3\leq 1$

Proof: From (2.13) follows $2\rho - R_1 = 6\rho/\pi^2$.
$$
\frac{3A}{B^3} = \frac{3\pi^2}{6B^3\rho} < \frac{5}{\rho B^3} <
\frac{1}{3},
$$
because $B > 3\rho^{-1/3}$. Also,
$$
\frac{2}{3\rho B} \leq \frac{1}{3}
$$
because $B > 2/\rho$.
\item
$U_n \leq nKB^n$ for $n = 1,\dots,6$.
Since $u_1 = u_3 = 0$ and $|u_4(t)| \leq M$, $|u_6(t)| \leq M$, and $B\geq
M\geq 1$ we need to
consider only $n =2, 5$. From the definitions of $u_2$,
$u_5$ we
get
$$
\align
\sup_{|t|=R_2}|u_2(t)| &= \sup_{|t|=R_2}|\psi^{'}(t)|\\
                       &\leq \frac{2M}{R_1-R_2} = \frac{4M\pi^2}{3\rho}\\
                       & < \frac{16M}{\rho} < 2KB^2, \\
\sup_{|t|=R_5}|u_5(t)| &= \sup_{|t|=R_5}\frac{|\psi''(t)|}{6}\\
                       &\leq \frac{M}{3(R_4-R_5)(2\rho-R_1)} \\
                       &= \frac{25M\pi^4}{108\rho^2} < 5KB^5.
\endalign
$$
\endroster
(Note, in the latter, we have used Cauchy's estimate twice on two concentric
disks.)
So we get  for all $n$, $U_n \leq nKB^n$ completing the induction.
The arithmetic-geometric series $\sum_{n=1}^{\infty} nK(XB)^n$ dominates the
series  $\sum_{j=1}^{\infty} |u_j(t)|X^j$ uniformly in the set
$$
\{(X,t); |X|<B^{-1}, |t| \leq \rho\} \subset \Bbb C^2.
$$
Since $M \geq 1$ and $\rho_0 := \min\{1, \rho\}\leq 1$,
 the required estimate (2.7) on $B$ holds if we take $B = 10M/\rho_0$,
 completing the proof of the theorem.    \qed
\enddemo
\remark{Remark} Note that the majorant $\sum_{n=1}^{\infty} nK(XB)^n$ has a
double
pole on its circle of convergence. This is consistent with the suspected
double pole of the solution as the next singularity away  from the initial
manifiold (1.1).

\subheading{\S 3 Proof of Theorem B --- Continuity With Respect To The
Arbitrary Functions}

In this section, we prove Theorem B.
Recall the notation defined for domains $\Cal O_\delta$, $\Cal O$
in the Introduction.
For emphasis we may sometimes use the notations
$\Cal O_{\delta}(\psi, u_4, u_6)$,
$B_{\delta}(\psi, u_4, u_6)$,
 $M_{\delta}(\psi, u_4, u_6)$,
or simply  $\Cal O_{\delta}(\psi), B_{\delta}(\psi),  M_{\delta}(\psi)$.

\proclaim{Lemma 3.1}
Given $\Om \subset \Bbb C$ containing the origin
and $\psi, u_4, u_6$ holomorphic in $\Om$
 the meromorphic solution obtained in Theorem [1.1]
exist in the domain $\Cal O$.
\endproclaim
\demo{Proof}
Pick a $\delta$ arbitrary and  small.  Let $t \in \Om_{\delta}$. Then the
 closed disk $\ob{D_{2\delta}}(t)$  is contained in $\Om_{\delta/3}$.
By maximum modulus theorem,
$$
\max_{\ob{D_{2\delta}(t)}}\{1, |\psi(\tau)|, |u_4(\tau)|, |u_6(\tau)|\} \leq
M_{\delta}.
$$
The proof of Theorem A can be generalised by replacing all suprema over
$|t|=R_i$ by suprema over $|t-a|=R_i$.  This shows that the solution is
defined on
$\{(x,\tau):
\tau\in D_{2\delta}(t), |x-\psi(\tau)| < B_{\delta}^{-1}\}$ and since the
$B_{\delta}$
is the same for all the disks $\ob{D_{2\delta}}(t)$ as $t$ varies over
$\Om_{\delta}$,
the solution is defined on
$$
\bigcup_{t\in\Om_{\delta}}\{(x,\tau): \tau\in \ob{D_{2\delta}}(t),
|x-\psi(\tau)| <
B_{\delta}^{-1}\}
\supset \Cal O_{\delta}.
$$
Since $\delta > 0$ is arbitrary, the solution is defined on
$\bigcup_{\delta > 0}\Cal O_{\delta} = \Cal O$.

This lemma is used below to enable us to work in any polydisk in
$\Cal O_{\delta}$.
\enddemo

\demo{Proof of Theorem}
\roster
\item
Proof of the first part of the theorem:

Since we have $\psi(0) = 0$ it follows that $\Cal O$ contains the origin.
Choose $\delta>0$ such that $0\in\Omega_\delta$ and choose a polydisk
$P(a_1, a_2)$ whose closure is contained in $\Cal O_\delta$.
Since $\psi_j(t)$, $u_{4, j}(t)$, $u_{6, j}(t)$ converge uniformly on
compact subsets of $\Om$, there exists a  $J = J_{\delta}$
such that
$$
1\leq \sup_j\sup_{\Om_{\delta/3}}\{1, |\psi_j(t)|, |u_{4, j}(t)|, |u_{6, j}(t)|
\}  \leq J_{\delta}
$$
Let $$H = \frac{10J_{\delta}}{\min\{1, \delta\}}.$$
There exists $j_0$ such that
$$
\sup_{\Om_{\delta}}\{|\psi_j(t) - \psi(t)|, |u_{4, j}(t) - u_4(t)|,
|u_{6, j}(t) - u_6(t)|\} < \frac{1}{3H} \quad \forall j \geq j_0
$$
Now choose a $b_2 > 0$ with $b_2 < a_2$ such that
$|\psi(t)| < \frac{1}{3H}$ for all
$|t| < b_2$ and now if $|x| < \frac{1}{3H}, |t| < b_2$,
$$
|x - \psi_j(t)| \leq |x| + |\psi(t)| + |\psi_j(t) - \psi(t)|
< 3\frac{1}{3H} = \frac{1}{H} \quad \forall  j\geq j_0
$$
But note that $$H = \frac{10J_{\delta}}{\min\{1, \delta\}} \geq
\frac{10M_{\delta}(\psi_j)}{\min\{1, \delta\}} =
B_{\delta}(\psi_j).$$
This shows that if we take $b_1 = \frac{1}{3H}$ and $b_2$ as above then
 $P(b_1, b_2) \subset \Cal O_{\delta}(\psi_j)$ for all $j\geq j_0$.
Finally, since $\cap_{j=1}^{j_0} \Cal O_{\delta}(\psi_j)$ is an
open set containing the origin it follows that there exists $c_1, c_2$
such that $P(c_1, c_2)\subset\bigcap_{j=1}^{j_0} \Cal O_{\delta}(\psi_j)$  and
 then $r_i = \min\{a_i,b_i, c_i\}$ $(i=1,2)$ suffices.
The lower bound on the radius of convergence given in Theorem~A and the
uniform convergence of the WTC data imply that $r_i\ne 0$.
This finishes the proof
of the first part of the theorem.
\item
Proof of the second part of the theorem:

To prove that the convergence is uniform on compact subsets we proceed as
follows. Let $C$ be a compact subset of $G:=P(r_1, r_2)$.
Choose a compact subset
$\widetilde
C$ of $G$ such that $C\Subset \widetilde C$.\footnotemark
\footnotetext{The notation $C \Subset \widetilde C$ means closure of $C$ is
compact
and is contained in the interior of $\widetilde C.$}
By the maximum modulus theorem we have
$$
\sup_{(x,t)\in C}|x - \psi(t)| < \beta < \gamma
< \sup_{(x,t)\in \widetilde C}|x - \psi(t)|
$$
for some pair of numbers $\beta$ and $\gamma$.  Uniform convergence of
$\psi_j$ gives a $j_0$ such that for all $j\geq j_0$,
$$
\sup_{(x,t)\in C}|x - \psi_j(t)| < \beta < \gamma
< \sup_{(x,t)\in \widetilde C}|x - \psi_j(t)|
$$
Since $\widetilde C$ is compact, there is a $\delta > 0$ such that $\widetilde
C
\subset \Cal O_{\delta}$.
(Note $\Cal O_{\delta}$ are nested domains.)
In $\Cal O_{\delta}(\psi_j)$ we have $|x - \psi_j(t)| <
B_{\delta}(\psi_j)^{-1}$ and
consequently $B_{\delta}(\psi_j)^{-1} > \gamma$ for all $j\geq j_0$.
Let $u_{n,j}$ be given by the recursion formula (1.6) with data
$\psi_j, u_{4, j}, u_{6, j}$. Then we have the estimates:
$$
\gather
\sup_{t\in \Om_{\delta}(\psi_j)}|u_{n,j}(t)| \leq nK(B_{\delta}(\psi_j))^n
                            < nK\gamma^{-n} \quad\forall j\geq j_0,\\
\therefore \sup_{(x,t)\in C}|u_{n,j}(t)||x - \psi_j(t)|^n  <
 nK\beta^n\gamma^{-n}   \quad\forall j\geq j_0.
\endgather
$$
Meanwhile by induction on $n$ and (1.6), it follows easily that
$$
\lim_{j\to\infty}u_{n,j}\rightarrow u_n, \quad
\lim_{j\to\infty}u_{n,j}^{'}\rightarrow u_n^{'}\quad\text{unif. comp. sets. }
$$
The remainder of the proof is  now a consequence of  the following result
which is a special case of Lebesgue dominated convergence theorem.
\endroster
\enddemo
\proclaim{A Dominated Convergence Theorem}
Suppose that $f: \Bbb N\times\Bbb N\times C \rightarrow \Bbb C$ such that

1) $\lim_{j\to\infty}f(n,j,v)$ converges uniformly in $v$ to a limit denoted
by $f(n, v)$.

2) $\sup_{j,v}|f(n,j,v)| \leq g(n)$ and

3) $\sum_{n=1}^{\infty} g(n)$ converges

Then,
$$
\lim_{j\to\infty}\sum_{n=1}^{\infty}\sup_{v}|f(n,j,v) - f(n, v)| = 0
$$
In particular $\sum_{n=1}^{\infty}f(n,j,v)$ converges to
$\sum_{n=1}^{\infty}f(n, v)$ uniformly on $C$.
\endproclaim
Apply the above to the case (with $v = (x, t)$)
$$
f(n, j, v) = u_{n, j}(x - \psi_j(t))^n, \quad f(n, v)
= u_n(x - \psi(t))^n, \quad g(n) = nK(\frac{\beta}{\gamma})^n
$$
This finishes the proof of the second part  of the theorem.  \qed

\subheading{\S 4 Conclusion}
In this paper we have obtained an explicit lower bound for the radius of
convergence of the Painlev\'e expansion of solutions of the KdV equation and
used this estimate to prove the continuous dependence of solutions on the
WTC data.
Note that these cannot be given in the usual way {\it i.e.} as one would give
Cauchy data on a regular manifold.  The question of how to relate WTC data to
regular Cauchy data elsewhere still remains open.
The methods used in this paper
readily extend to other integrable PDEs
that are analytic in the dependent variable $u$ and its derivatives,
including those in $2+1$-dimensions
such as the Kadomtsev-Petviashvili equation.

A proof that the KdV
equation possesses the Painlev\'e property is still lacking in the literature.
The major problem is the absence so far of a method of global
analysis in ${\Bbb C}^n$ that applies to the whole space of solutions.

\proclaim{Acknowledgements} This research was supported by the Australian
Research Council.  The authors would also like to thank Rod Halburd and the
referees, particularly the first referee, for a careful reading of the paper
and detailed suggestions for improvement.
\endproclaim

\end
\Refs
\ref \no 1\by M.J.Ablowitz  and P.A.Clarkson
\book Solitons, Nonlinear Evolution Equations and Inverse Scattering
\yr1991 \bookinfo London Math. Soc. Lect. Note. Ser 149
\publ Cambridge University Press\endref

\ref \no 2\by M.J.Ablowitz, A.Ramani and H.Segur
\paper Nonlinear Evolution equations and Ordinary Differential Equations of
Painlev\'e Type
\yr 1978
\jour Lett. Nuovo Cimento \vol 23 \pages 333-338
\endref

\ref \no 3
\by M.J.Ablowitz, A.Ramani and H.Segur
\paper A connection Between Nonlinear Evolution Equations and Ordinary
Differential Equations of P-Type I, II
\jour Jour. Math. Phys. \vol 21 \pages 715-721, 1006-1015 \yr 1980
\endref

\ref \no 4\by Joshi N and Petersen J. A
\paper A Method of Proving the Convergence of the Painlev\'e Expansions of PDEs
\jour Nonlinearity \vol 7 \pages 595-602 \yr 1994\endref

\ref \no 5 \by Joshi N and Kruskal M
\paper A direct Proof that Solutions of the Six Painlev\'e Equations Have no
Movable Singularities Except Poles
\jour Stud. In Appl. Math \yr 1994 \vol 93\pages 187-207\endref

\ref \no 6\by Kichenassamy S and Littman W
\paper Blow-up Surfaces for Nonlinear Wave Equations-I, II
\jour Commun. PDE \vol 18 \pages 431-52, 1869-99 \yr 1993\endref

\ref \no 7\by Weiss J, Tabor M  and Carnevale G
\paper The Painlev\'e Property for Partial Differential Equations
\jour J. Math. Phys \vol 24 \pages 522-6 \yr 1983\endref
